\documentclass[aps,prb,amsmath,twocolumn,superscriptaddress]{revtex4}

\usepackage{amsmath}
\usepackage{amssymb}

\usepackage{graphicx}
\usepackage{bm} 
\usepackage{url}

\DeclareMathOperator{\sgn}{sgn}
\DeclareMathOperator{\Img}{\mathrm{Im}}

\DeclareMathOperator{\Sp}{\mathrm{Sp}}

\begin{document}

\author{Mikhail S. Kalenkov}
\affiliation{I.E. Tamm Department of Theoretical Physics, P.N. Lebedev Physical
Institute, 119991 Moscow, Russia}
\affiliation{Moscow Institute of Physics and Technology, Dolgoprudny, 141700
Moscow region, Russia}
\author{Andrei D. Zaikin}
\affiliation{Institut f\"ur Nanotechnologie, Karlsruher Institut f\"ur Technologie (KIT), 76021 Karlsruhe, Germany}
\affiliation{I.E. Tamm Department of Theoretical Physics, P.N. Lebedev Physical Institute, 119991 Moscow, Russia}

\title{Large thermoelectric effect in ballistic Andreev interferometers}

\begin{abstract}
Employing quasiclassical theory of superconductivity combined with Keldysh technique we investigate large thermoelectric effect in multiterminal ballistic normal-superconducting (NS) hybrid structures. We argue that this effect is caused by electron-hole asymmetry generated by coherent Andreev reflection of quasiparticles at interfaces of two
different superconductors with non-zero phase difference. Within our model we derive a general expression
for thermoelectric voltages $V_{T1,2}$ induced in two different normal terminals exposed to
a thermal gradient. Our results apply at any temperature difference in the subgap regime and allow to
explicitly analyze both temperature and phase dependencies of $V_{T1,2}$ demonstrating that in general there exists no fundamental relation between these voltages and the equilibrium Josephson current in SNS junctions.
\end{abstract}

\pacs{PACS}

\maketitle

\section{Introduction}
Thermoelectric effect (i.e. the appearance of an electric current upon application of a thermal gradient) in superconductors remains an intriguing topic that attracts a lot of interest over last decades \cite{NL}.
While usually the magnitude of this effect in generic metals and superconductors remains small (being proportional to the ratio between temperature and the Fermi energy $T/\varepsilon_F$), it can increase by orders of magnitude provided electron-hole symmetry is lifted, e.g., due to spin-dependent scattering of electrons. This situation was predicted to occur in a variety of structures, such as superconductors doped with magnetic impurities \cite{Kalenkov12}, superconductor-ferromagnet hybrids with the density of states spin-split by the exchange and/or Zeeman fields \cite{Machon,Ozaeta} or superconductor-normal metal (SN) bilayers with spin-active interfaces \cite{KZ14,KZ15}. In accordance with theoretical predictions large thermoelectric currents were recently observed in superconductor-ferromagnet tunnel junctions in high magnetic fields \cite{Beckmann}.

Large thermoelectric effect was also observed in multiterminal hybrid SNS structures with no magnetic inclusions \cite{Venkat1,Venkat2,Petrashov03,Venkat3}. Being exposed to a temperature gradient such structures (frequently called Andreev interferometers) were found to develop a thermopower signal which magnitude was not restricted by a small parameter $T/\varepsilon_F$ and, furthermore, turned out to be a periodic function of the superconducting phase difference $\chi$ across the corresponding SNS junction. The latter observation indicates that macroscopic quantum coherence can play an important role and poses a question about the relation between thermoelectric and Josephson effects in the systems under consideration. Subsequent theoretical analysis \cite{Seviour00,KPV,VH,VP} indeed demonstrated that the thermoelectric effect in Andreev interferometers can be large and confirmed the periodic dependence of the thermopower on the phase difference $\chi$. At the same time, there appear to be no general consensus in the literature concerning the basic physical origin of this effect. While the authors \cite{KPV} emphasized an important role of electron-hole imbalance, Virtanen and Heikkil\"a \cite{VH}, on the contrary, proposed that in Andreev interferometers a non-vanishing thermopower signal can be generated even provided electron-hole symmetry is maintained and that the dominant part of this signal can be directly related to the difference between the {\it equilibrium} values for the Josephson current at temperatures $T_1$ and $T_2$. Subsequently, Volkov and Pavlovskii \cite{VP} argued that in general no such simple relation between the thermopower and the Josephson current can be established and that under certain conditions the former can still remain large even though the latter gets strongly suppressed by temperature effects.

The existing theory \cite{Seviour00,KPV,VH,VP} merely deals with the experimentally relevant diffusive limit in which case the analysis may become rather cumbersome forcing the authors either to employ numerics or to resort to various approximations. On the other hand, in order to clarify the physical origin of a large thermoelectric effect in Andreev interferometers one could take a different route focusing the analysis on the ballistic limit. A clear advantage of this approach is the possibility to employ the notion of semiclassical electron trajectories and in this way to treat the problem exactly. Below we will follow this route.

The structure of the paper is as follows. In Sec. \ref{model} we define our
model and describe the quasiclassical formalism which will be employed in our
further analysis. In Sec. \ref{riccati} we demonstrate that the quasiclassical
Eilenberger equations supplemented by Zaitsev boundary conditions can be conveniently resolved with the aid
of the so-called Riccati parameterization of the Green functions. Specific
quasiparticle trajectories and their contributions to electric currents are
identified in Sec. \ref{quas} where we also discuss the conditions for
electron-hole symmetry violation in our system. In Sec. \ref{thermo} we
evaluate the thermoelectric voltages induced by applying a temperature gradient
and demonstrate that these voltages can be large because of the presence of
electron-hole asymmetry in Andreev interferometers. Sec. \ref{josephson} is
devoted to further analysis of the relation between thermoelectric and Josephson
effects. Some technical details of our calculation are displayed in Appendix.

\section{The model and quasiclassical formalism}
\label{model}
\begin{figure}
\centerline{ \includegraphics[width=80mm]{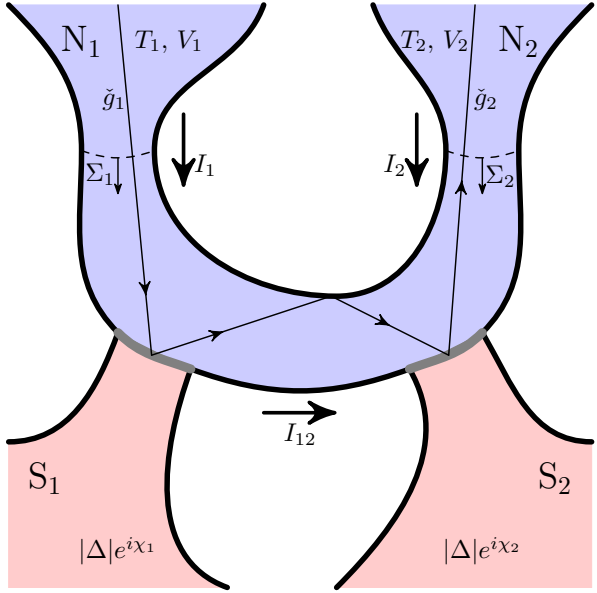} }
\caption{(Color online) Normal wire connected to normal and superconducting leads. Normal leads are maintained at temperatures $T_1$ and $T_2$ and voltages $V_1$ and $V_2$ respectively. An example of a quasiclassical electron trajectory relevant for the thermoelectric effect under consideration is also illustrated.}
\label{nsnsn3-fig}
\end{figure}

Let us consider an NSNSN structure shown in Fig. \ref{nsnsn3-fig}. Two
superconductors with phase difference $\chi = \chi_1 - \chi_2$ are connected to
a normal wire. The two ends of this normal wire are maintained at temperatures $T_1$
and $T_2$ and voltages $V_1$ and $V_2$ respectively. Within our model we deliberately choose to disregard electron scattering on impurities and boundary imperfections, in which case electron motion is ballistic and
can be conveniently described in terms of quasiclassical trajectories. In
addition, we will disregard inelastic electron relaxation inside our system by
assuming the inelastic relaxation length to exceed the system size. In this case
inelastic electron relaxation can only occur deep inside normal terminals
$\mathrm{N_1}$ and $\mathrm{N_2}$.

In what follows we will employ the quasiclassical theory of superconductivity based on the Eilenberger equations \cite{Eil,bel}. Under the above assumptions adopted within our model these equations take the form
\begin{equation}
\left[ \hat\Omega , \hat g^{R,A,K} \right]
+
i\bm{v}_F \nabla \hat g^{R,A,K} (\bm{p}_F, \bm{r}, \varepsilon) =0,
\quad
\check g^2 =1,
\end{equation}
where $\hat g^{R,A,K}$ are energy-integrated retarded, advanced and Keldysh $2\times 2$ matrix
Green functions. The matrices $\hat\Omega$ and $\check g$ have the following
structure
\begin{equation}
\hat \Omega=
\begin{pmatrix}
\varepsilon & \Delta \\
-\Delta^* & -\varepsilon
\end{pmatrix},
\quad
\check g=
\begin{pmatrix}
\hat g^R & \hat g^K  \\
0 & \hat g^A
\end{pmatrix},
\end{equation}
where $\Delta$ and $\varepsilon$ are respectively the superconducting order parameter and the
quasiparticle energy. Electric current density can be expressed in terms of the
Keldysh Green function in the standard manner as
\begin{equation}
\bm{j}(\bm{r})= -\dfrac{e N_0}{4} \int d \varepsilon
\left< \bm{v}_F \Sp [\hat \tau_3 \hat g^K(\bm{p}_F, \bm{r},
\varepsilon) ] \right>,
\label{current}
\end{equation}
where $\bm{p}_F=m\bm{v}_F$ is the electron Fermi momentum vector, $\hat\tau_3$
is the Pauli matrix in the Nambu space and  $N_0=mp_F/(2\pi^2)$ is the normal density of states at
the Fermi level. Here the angular brackets $\left<\cdots\right>$ denote
averaging over
the Fermi momentum directions.

\section{Riccati parameterization and boundary conditions}
\label{riccati}
For the system under consideration the Eilenberger equations can be solved exactly.
In order to proceed we restrict our analysis to quasiparticles propagating in the normal metal with subgap energies $|\varepsilon| < |\Delta|$ and employ the so-called Riccati parameterization for the retarded and advanced Green functions \cite{Schopohl95}
\begin{equation}
\hat g^{R,A}=\pm
    \hat N^{R,A}
    \begin{pmatrix}
    1+\gamma^{R,A} \tilde \gamma^{R,A} & 2\gamma^{R,A} \\
    -2 \tilde \gamma^{R,A} & -1- \tilde \gamma^{R,A}  \gamma^{R,A} \\
    \end{pmatrix},
    \label{graparam}
\end{equation}
where $\gamma^{R,A}$, $\tilde \gamma^{R,A}$ are Riccati amplitudes and
\begin{equation}
\hat N^{R,A}=
    \begin{pmatrix}
    (1-\gamma^{R,A} \tilde \gamma^{R,A})^{-1} & 0 \\
    0 & (1-\tilde \gamma^{R,A}  \gamma^{R,A} )^{-1} \\
    \end{pmatrix}.
    \label{nrparam}
\end{equation}
Parameterization of the Keldysh Green function $\hat g^K$ also contains the two distribution functions  $x$ and $\tilde x$. It reads \cite{Eschrig00}
\begin{equation}
\hat g^K=
2
\hat N^R
\begin{pmatrix}
x - \gamma^R  \tilde x  \tilde \gamma^A &
-\gamma^R  \tilde x + x  \gamma^A \\
-\tilde \gamma^R  x + \tilde x \tilde \gamma^A &
\tilde x - \tilde \gamma^R  x  \gamma^A \\
\end{pmatrix}
\hat N^A.
\label{gkparam}
\end{equation}

In the normal metal (i.e. for $\Delta \equiv 0$) Riccati amplitudes $\gamma^{R,A}$, $\tilde \gamma^{R,A}$ and distribution functions  $x$, $\tilde x$ obey the following simple equations
\begin{gather}
i\bm{v}_F \nabla \gamma^{R,A} = -2 \varepsilon \gamma^{R,A},
\quad
i\bm{v}_F \nabla \tilde \gamma^{R,A} = 2 \varepsilon \tilde \gamma^{R,A},
\label{gammaeq}
\\
i\bm{v}_F \nabla x =0, \quad i\bm{v}_F \nabla \tilde x =0.
\label{xeq}
\end{gather}
Within the quasiclassical approximation adopted here quasiparticles propagate
along the straight line trajectories between each two scattering events which
can only occur at the boundaries of the normal metal wire. Of interest for us
here is to describe quasiparticle scattering at the interfaces between the
normal metal and each of the two superconductors. This task is accomplished in
the standard manner with the aid of the Zaitsev boundary conditions
\cite{Zaitsev84} for the quasiclassical Green functions rewritten in terms of
the above Riccati amplitudes and the distribution functions\cite{Eschrig00}.

Consider, for instance, a quasiparticle propagating from the first normal terminal, being reflected at each of the two NS interfaces and leaving the system through the second normal terminal. The corresponding quasiparticle trajectory is indicated in Fig. \ref{nsnsn3-fig}. It is important to emphasize that here we only account for quasiparticles with subgap energies which cannot penetrate into superconductors suffering either normal or Andreev reflection at both NS interfaces. It is easy to verify that the distribution functions for such quasiparticles take the form
\begin{gather}
x =
\left[1- \gamma^R \tilde \gamma^A\right] x_1,
\quad
\tilde x =
\left[1- \tilde \gamma^R \gamma^A \right] \tilde x_2
\end{gather}
obeying both Eilenberger equations and the corresponding boundary conditions at NS
interfaces. Here $\gamma^{R,A}$ and $\tilde \gamma^{R,A}$ are Riccati amplitudes
along the trajectory, $x_1$ and $\tilde x_2$ are asymptotic values of the
distribution functions $x$ and $\tilde x$ respectively at the initial and final trajectory points. Then we obtain
\begin{equation}
\hat g^K = \dfrac{x_1}{2}
(1+\hat g^R)(1 - \hat g^A)
+
\dfrac{\tilde x_2}{2}
(1 - \hat g^R)(1 + \hat g^A).
\end{equation}
Making use of the asymptotic conditions
\begin{gather}
\gamma^R_1 = \tilde \gamma^A_1 =0, \quad \tilde \gamma^R_2 = \gamma^A_2 = 0,
\label{as1}
\end{gather}
which hold respectively in the initial and final
trajectory points we recover simple expressions for the Keldysh Green function
inside the normal leads, i.e.
\begin{gather}
\hat g^K_1=
2
\begin{pmatrix}
x_1  &  x_1  \gamma^A_1 \\
-\tilde \gamma^R_1  x_1 & \tilde x_2 - \tilde \gamma^R_1 \gamma^A_1 (x_1 + \tilde x_2)
\\
\end{pmatrix}
,
\\
\hat g^K_2=
2
\begin{pmatrix}
x_1 - \gamma^R_2 \tilde \gamma^A_2 ( x_1 + \tilde x_2) & - \gamma^R_2  \tilde x_2 \\
\tilde x_2 \tilde \gamma^A_2 & \tilde x_2\\
\end{pmatrix}.
\end{gather}

The expressions for Riccati amplitudes $\gamma^{R,A}$ and $\tilde \gamma^{R,A}$ are derived from Eqs. \eqref{gammaeq}  supplemented by the boundary conditions at the NS interfaces. The latter can be expressed in the form \cite{Eschrig00}
\begin{gather}
\gamma^R_{\text{out}}=
\dfrac{
(\mathcal{R}-\gamma^R_S \tilde \gamma^R_S) \gamma^R_{\text{in}} + \mathcal{D} \gamma^R_S
}{
- \mathcal{D} \tilde \gamma^R_S \gamma^R_{\text{in}} +(1- \mathcal{R}\gamma^R_S \tilde \gamma^R_S)
},
\label{bound1}
\\
\gamma^A_{\text{in}}=
\dfrac{
(\mathcal{R}-\gamma^A_S \tilde \gamma^A_S) \gamma^A_{\text{out}} + \mathcal{D} \gamma^A_S
}{
- \mathcal{D} \tilde \gamma^A_S \gamma^A_{\text{out}} +(1- \mathcal{R}\gamma^A_S \tilde \gamma^A_S)
},
\label{bound2}
\\
\tilde \gamma^R_{\text{in}}=
\dfrac{
(\mathcal{R}-\gamma^R_S \tilde \gamma^R_S) \tilde \gamma^R_{\text{out}} + \mathcal{D} \tilde
\gamma^R_S
}{
- \mathcal{D} \gamma^R_S \tilde \gamma^R_{\text{out}} +(1- \mathcal{R} \gamma^R_S \tilde
\gamma^R_S)
},
\label{bound3}
\\
\tilde \gamma^A_{\text{out}}=
\dfrac{(\mathcal{R}-\gamma^A_S \tilde \gamma^A_S) \tilde \gamma^A_{\text{in}}
+
\mathcal{D} \tilde \gamma^A_S}{
- \mathcal{D} \gamma^A_S \tilde \gamma^A_{\text{in}}
+(1- \mathcal{R}\gamma^A_S \tilde \gamma^A_S)
\label{bound4}
},
\end{gather}
where the Riccati amplitudes denoted by the subscripts ``in'' and ``out''
parameterize retarded and advanced Green function for respectively incoming and outgoing
momentum directions, $\mathcal{D}=1-\mathcal{R}$ denotes the
normal transmission of the corresponding NS interface and $\gamma^{R,A}_S$, $\tilde \gamma^{R,A}_S$ are
the Riccati amplitudes in the superconductor defined as
\begin{gather}
\gamma_S^R = \dfrac{\varepsilon - \sqrt{\varepsilon^2 -
|\Delta|^2}}{\Delta^*},
\quad
\tilde \gamma_S^A = [\gamma_S^R]^*,
\\
\tilde \gamma_S^R = \dfrac{\varepsilon - \sqrt{\varepsilon^2 -
|\Delta|^2}}{\Delta},
\quad
\gamma_S^A = [\tilde \gamma_S^R]^*,
\end{gather}
where the branch of the square root is chosen so that $\sgn \Img \sqrt{z} =
\sgn \Img z$.

From Eqs. \eqref{bound1}-\eqref{bound4} it is easy to observe that at subgap energies $|\varepsilon| < |\Delta|$  the boundary conditions at the NS interface acquire the same form for four different functions $\gamma^R$, $1/\tilde \gamma^R$, $1/\tilde \gamma^A$, and $\gamma^A$, i.e.
\begin{gather}
w_{\text{out}}=
\dfrac{
(\mathcal{R}-\gamma^R_S \tilde \gamma^R_S) w_{\text{in}} + \mathcal{D} \gamma^R_S
}{
- \mathcal{D} \tilde \gamma^R_S w_{\text{in}} +(1- \mathcal{R}\gamma^R_S \tilde \gamma^R_S)
},
\\
w=\gamma^R, 1/\tilde \gamma^R, 1/\tilde \gamma^A, \gamma^A.
\label{w}
\end{gather}
Here we employed the identities $\gamma_S^R \tilde \gamma_S^A =1$ and $\tilde
\gamma_S^R \gamma_S^A =1$ which hold in the relevant energy interval
$|\varepsilon| < |\Delta|$. Furthermore, Eqs. \eqref{gammaeq} demonstrate that
the four functions \eqref{w} obey the same equation. With this in mind it is
straightforward to verify that the following combination of the Riccati
amplitudes
\begin{equation}
\dfrac{1 - \gamma^R \tilde \gamma^A}{1 - \gamma^R \tilde \gamma^R}
\dfrac{1 - \tilde \gamma^R \gamma^A}{1 - \gamma^A \tilde \gamma^A}
\end{equation}
remains constant along the trajectory at subgap energies. Then making use of Eqs. \eqref{as1} we obtain the relation between Riccati amplitudes in the initial and final points of the quasiclassical trajectory:
\begin{equation}
\tilde \gamma^R_1 \gamma^A_1
=
\gamma^R_2 \tilde \gamma^A_2.
\end{equation}

Asymptotic behavior of Riccati amplitudes at the beginning and at the end of
electron trajectories is
directly related to transmission and reflection probabilities of the corresponding
processes. These probabilities will be explicitly evaluated in the next section.

\section{Quasiclassical trajectories and electron-hole asymmetry generation}
\label{quas}
As we already pointed out, quasiparticles with subgap energies can only propagate inside the normal part of our system being unable to penetrate deep into the superconductors S$_1$ and S$_2$. Let us classify such electron trajectories relevant for the thermoelectric effect under consideration.

There exist electron trajectories starting in one of the terminals
($\mathrm{N_1}$ or $\mathrm{N_2}$) and going back to the same terminal without
hitting any
of the two NS interfaces. These trajectories do not contribute to any current flowing in our system and, hence, can be safely ignored in our subsequent
consideration. Of more relevance are electron trajectories which propagate from the first to the second terminal (or vice versa).
Provided these trajectories ``know nothing about superconductivity'' (even 
though some of them can hit at least one of the NS interfaces)
they contribute to the dissipative Ohmic current $(V_1-V_2)/R_{0}$ flowing
between the terminals $\mathrm{N_1}$ and $\mathrm{N_2}$. Here
\begin{equation}
\dfrac{1}{R_{0}}=
2 e^2 N_0 \int \left< \bm{v}_F \Theta_{12}(\bm{p}_F,\bm{r}) \right> d
\bm{\Sigma}_1,
\label{sharvin}
\end{equation}
is the inverse Sharvin resistance of the normal wire and the function $\Theta_{12}(\bm{p}_F,\bm{r})$ equals to unity for
all electron trajectories connecting the first and the second terminals and to zero otherwise.
Here and below averaging $\left< ...\right>$ includes only the directions of $\bm{v}_F$ corresponding to electron trajectories going
out of the first terminal and $\int \cdots d\bm{\Sigma}_{1(2)}$ denotes the integral over the cross-section of the normal lead $N_{1(2)}$.

The remaining contributions to the currents $I_1$ and $I_2$ in the normal terminals (see Fig. \ref{nsnsn3-fig})
are due to electron trajectories which directly involve at least one of the superconductors. In what follows for the sake of simplicity we will restrict our analysis to trajectories which may hit each of the NS interfaces only once assuming that the contribution of more complicated trajectories is negligible. This can easily be achieved by a proper choice of the system geometry (e.g., by assuming the cross sections of both NS interfaces to be sufficiently small). There are electron trajectories which originate in the first terminal, hit one of the
NS interfaces and go back to the same terminal. Making use of the formalism
described in the previous sections we evaluate the Keldysh Green function on
such trajectories and then derive the expression for the current with the aid of
Eq. \eqref{current}. E.g., in the case of the first terminal we obtain
\begin{multline}
I_{1}^{loc}=
eN_0
\int
\left<
\bm{v}_{F}\Theta_{11}^S(\bm{p}_F,\bm{r})
|\tilde \gamma_1^R(\bm{p}_F,  \bm{r}, \varepsilon)|^2
\right>
\\\times
\left[
-\tilde x_1 (\varepsilon)
-
x_1(\varepsilon)
\right] d \varepsilon d \bm{\Sigma}_1,
\label{i1loc}
\end{multline}
where the function $\Theta_{11}^S(\bm{p}_F,\bm{r})=1$ for electron trajectories which start inside the
first terminal, hit one of the two NS interfaces and return back to the same
terminal $\mathrm{N_1}$ and $\Theta_{11}^S(\bm{p}_F,\bm{r})=0$ otherwise. The
expression for $I_{2}^{loc}$ can be obtained from Eq. \eqref{i1loc} simply by
replacing the indices $1 \leftrightarrow 2$.
We also note that the equilibrium distribution functions $x_{1,2}$ and $\tilde
x_{1,2}$ in the bulk normal electrodes $\mathrm{N_1}$ and $\mathrm{N_2}$ are
defined by the standard expressions
\begin{gather}
x_{1,2}=\tanh\dfrac{\varepsilon - eV_{1,2}}{T_{1,2}},
\quad
\tilde x_{1,2}=-\tanh\dfrac{\varepsilon + eV_{1,2}}{T_{1,2}}.
\label{distr}
\end{gather}

What remains is to account for the trajectories connecting two different
terminals and touching either only one of the two NS interfaces or both these
interfaces one after the other. The latter situation is illustrated in Fig.
\ref{nsnsn3-fig}. As before, for these trajectories we set
$\Theta_{12}^S(\bm{p}_F,\bm{r})=1$, whereas $\Theta_{12}^S(\bm{p}_F,\bm{r})=0$
for all other trajectories. Again evaluating the Keldysh component of the Green
function matrix and making use of Eq. \eqref{current}, we get
\begin{multline}
I^{nl} = eN_0
\int
\left<
\bm{v}_{F}\Theta_{12}^S(\bm{p}_F,\bm{r})
|\tilde \gamma_1(\bm{p}_F,  \bm{r}, \varepsilon)|^2
\right>
\\\times
\left[
-\tilde x_2 (\varepsilon)
-
x_1(\varepsilon)
\right] d \varepsilon d \bm{\Sigma}_1.
\label{nl}
\end{multline}

Collecting all the above contributions, we determine the currents $I_1$ and $I_2$ flowing into respectively the first and the second normal terminals:
\begin{gather}
I_1 = (V_1-V_2)/R_{0}+I_1^{loc}(V_1)+I^{nl}(V_1,V_2,\chi),
\label{i1}
\\
I_2 =(V_2-V_1)/R_{0}+ I_2^{loc}(V_2) +I^{nl}(V_1,V_2,\chi).
\label{i2}
\end{gather}

In order to proceed we need to evaluate the Riccati amplitude $\tilde \gamma_1^R(\bm{p}_F,
\bm{r}, \varepsilon)$ in the beginning of the corresponding trajectory. For simplicity let us assume that both temperatures $T_{1,2}$ and voltages $V_{1,2}$ remain well below the superconducting gap, i.e. $T_{1,2}, eV_{1,2} \ll |\Delta|$. In this case it follows immediately, e.g., from Eqs. \eqref{i1loc}-\eqref{nl} that electron transport in our system is dominated by quasiparticles with energies well in the subgap range
\begin{equation}
|\varepsilon| \ll |\Delta|.
\label{subgap}
\end{equation}

Consider first the quasiclassical electron trajectory that begins in the
terminal $\mathrm{N_1}$,
hits one of the NS interfaces and returns back to the same terminal. Making use of the analysis
developed in the previous section, under the condition \eqref{subgap} one readily finds
\begin{equation}
|\tilde \gamma_1^R(\bm{p}_F,  \bm{r},
\varepsilon)|^2 = \mathcal{D}^2/(1+\mathcal{R})^2,
\label{BTK}
\end{equation}
where, as before, $\mathcal{D}=\mathcal{D}_{1,2}=1-\mathcal{R}_{1,2}$ is the normal transmission of the corresponding NS interface. Likewise, for the trajectories which start in the first terminal, hit
the interfaces NS$_1$ and NS$_2$  and then go towards the terminal
$\mathrm{N_2}$ (as illustrated
in Fig. \ref{nsnsn3-fig}), in the limit \eqref{subgap} we obtain
\begin{multline}
|\tilde \gamma_1^R|^2 =
1-\frac{16\mathcal{R}_1\mathcal{R}_2}
{\bigl|
(1+\mathcal{R}_1)(1+\mathcal{R}_2)
+\mathcal{D}_1\mathcal{D}_2e^{i(\chi + 2\varepsilon d/v_F)}
\bigr|^{2}},
\label{as2}
\end{multline}
where $d$ is the effective distance covered by a quasiparticle between the two
scattering events at $\mathrm{NS_1}$ and $\mathrm{NS_2}$ interfaces. Here and
below we assume that this distance obeys the condition $d \gg v_F/|\Delta|$.
Combining the results \eqref{BTK} and \eqref{as2} with Eqs.
\eqref{i1loc}-\eqref{nl} we can easily evaluate the currents $I_{1,2}$
\eqref{i1}, \eqref{i2}.

Before we complete this calculation let us briefly discuss the physical meaning of the above results.
It is straightforward to observe that the asymptotic value $\tilde \gamma^R_1
\gamma^A_1=|\tilde \gamma_1^R (\varepsilon )|^2$ in the beginning of the
corresponding trajectory defines the Andreev reflection probability, i.e.
the probability for an incoming electron with energy $\varepsilon$ to be
reflected back as a hole. This observation is well illustrated, e.g., by Eq.
\eqref{BTK} which is nothing but the standard BTK result \cite{BTK}. Making
use of general symmetry relations for the Green functions one can also
demonstrate that the probability for an incoming hole to be reflected back as an
electron equals
to $|\tilde \gamma_1^R(-\varepsilon)|^2$. Thus, from Eq. \eqref{as2} we conclude that scattering on two NS interfaces generates {\it electron-hole symmetry violation}
\begin{equation}
|\tilde \gamma_1^R(\varepsilon)|^2\neq |\tilde \gamma_1^R(-\varepsilon)|^2
\end{equation}
for quasiparticles propagating from $\mathrm{N_1}$- to $\mathrm{N_2}$-terminals
along the trajectories displayed in Fig. \ref{nsnsn3-fig} provided the
superconducting phase difference $\chi$ takes an arbitrary value not equal to
zero or $\pi$ and provided normal transmissions of both NS interfaces obey the
condition $0<\mathcal{D}_{1,2}<1$. Below we will demonstrate that this
electron-hole asymmetry yields a large thermoelectric effect in the system
under consideration.

\section{Thermoelectric voltage}
\label{thermo}
Let us now evaluate the currents $I_1$ \eqref{i1} and $I_2$ \eqref{i2}. In order to recover the local BTK terms $I^{loc}_{1,2}$ we explicitly specify the contributions from electron trajectories scattered at the first
and the second NS interfaces by splitting $\Theta_{11}^S \to \Theta_{11}^{S_1}+\Theta_{11}^{S_2}$ (and similarly for $\Theta_{22}^S$). Then combining Eqs. \eqref{i1loc}, \eqref{distr} with \eqref{BTK}, at low voltages and temperatures $eV_{1,2}, T_{1,2} \ll |\Delta|$ we get
\begin{equation}
I^{loc}_{1,2} = V_{1,2}/R_{1,2},
\end{equation}
where $R_{1,2}$ define the standard BTK low temperature resistance of the SN interfaces, i.e.
\begin{multline}
\dfrac{1}{R_1}=
4e^2N_0
\int
\Biggr<
\bm{v}_{F}
\Biggl[
\dfrac{\mathcal{D}_1^2}{(1+\mathcal{R}_1)^2}
\Theta_{11}^{S_1}(\bm{p}_F,\bm{r})
+\\+
\dfrac{\mathcal{D}_2^2}{(1+\mathcal{R}_2)^2}
\Theta_{11}^{S_2}(\bm{p}_F,\bm{r})
\Biggr]
\Biggl>
d \bm{\Sigma}_1.
\label{BTK2}
\end{multline}
Note that for simplicity in Eq. \eqref{BTK2} we disregard multiple scattering effects \cite{FN0} and account for quasiparticle trajectories which hit only one of the two NS interfaces ignoring, e.g., the contribution of trajectories of the type $\Theta_{11}^{S_1S_2}(\bm{p}_F,\bm{r})$ which hit both NS interfaces. If necessary, the latter contribution can easily be recovered, however, it may only yield renormalization of subgap resistances $R_{1,2}$ (cf., e.g., the last term in Eq. \eqref{rnl} below) and does not play any significant role in our further analysis.

In contrast, the trajectories of the type $\Theta_{12}^{S_1S_2}(\bm{p}_F,\bm{r})$ displayed in Fig. \ref{nsnsn3-fig} give an important contribution to the non-local current $I^{nl}$ and they should
necessarily be accounted for along with trajectories $\Theta_{12}^{S_1}$ and $\Theta_{12}^{S_2}$.
Collecting all these contributions \cite{FN}, with the aid of Eqs. \eqref{distr}, \eqref{nl} and \eqref{as2}
we obtain
\begin{equation}
I^{nl}= (V_1+V_2)/R^{nl}_+
+
\tilde I^{nl}(T_{1,2},V_{1,2},\chi),
\end{equation}
where we defined
\begin{multline}
\dfrac{1}{R^{nl}_{\pm}}
=
2e^2N_0
\int
\Biggr<
\bm{v}_{F}
\Biggl[
\dfrac{\mathcal{D}_2^2\Theta_{12}^{S_2}(\bm{p}_F,\bm{r})}{(1+\mathcal{R}_2)^2}
\pm\dfrac{\mathcal{D}_1^2\Theta_{12}^{S_1}(\bm{p}_F,\bm{r})}{(1+\mathcal{R}_1)^2}
\\+
\dfrac{\mathcal{R}_1 \mathcal{D}_2^2 \pm \mathcal{R}_2 \mathcal{D}_1^2}{
(1+ \mathcal{R}_1 \mathcal{R}_2) (\mathcal{R}_1 + \mathcal{R}_2)}
\Theta_{12}^{S_1S_2}(\bm{p}_F,\bm{r})
\Biggr]
\Biggl>
d \bm{\Sigma}_1,
\label{rnl}
\end{multline}
and $\tilde I^{nl} (T_{1,2}, V_{1,2}, \chi)$ represents the term sensitive to electron-hole asymmetry in our system.
Performing the corresponding energy integral (see Appendix), we get
\begin{multline}
\tilde I^{nl}= -eN_0
\int
\Biggr<
\dfrac{32\pi\bm{v}_{F}\Theta_{12}^{S_1S_2}(\bm{p}_F,\bm{r})\mathcal{R}_1
\mathcal{R}_2\beta}{
(1+ \mathcal{R}_1 \mathcal{R}_2) (\mathcal{R}_1 + \mathcal{R}_2)
}
\\\times
\Biggl[
T_2 W(\beta, t_2, \chi - v_2)-T_1 W(\beta, t_1, \chi + v_1)
\Biggr]
\Biggl>
d \bm{\Sigma}_1,
\label{tnl}
\end{multline}
where
\begin{equation}
W(\beta,t,\chi) = \Img
\sum_{n \geqslant 0}
\dfrac{e^{i\chi} e^{-t(2n+1)}}{1 + \beta e^{i\chi} e^{-t(2n+1)}}
\label{W}
\end{equation}
and we defined
\begin{equation}
\beta = \dfrac{\mathcal{D}_1 \mathcal{D}_2}{(1+\mathcal{R}_1)(1+\mathcal{R}_2)},
\end{equation}
$t_{1,2} =2\pi T_{1,2} d/v_F$ and $v_{1,2}=2eV_{1,2} d/v_F$. Note that here and below
the parameter $d$ depends on the particular electron trajectory and, hence, the function
$W$ cannot be taken out of the angular brackets indicating averaging over the directions of $\bm{v}_{F}$.

The expression for $W$ \eqref{W} gets simplified in the limits of high and low temperatures,
i.e. \begin{equation}
W(\beta, t, \chi)
=
\begin{cases}
e^{-t}\sin \chi , & t \gg 1,
\\
\dfrac{1}{2t\beta}
\arctan
\left(
\dfrac{\beta \sin \chi}{1 + \beta \cos \chi}\right), & t \ll 1.
\end{cases}
\end{equation}

The current $\tilde I^{nl}$ is responsible for the large thermoelectric effect in the system under consideration. In order to illustrate this fact let us disconnect both normal terminals from external leads in which case one obviously has
\begin{equation}
I_1=I_2\equiv 0.
\label{discon}
\end{equation}
Then in the absence of a temperature gradient (i.e. for $T_1=T_2$) both voltages vanish identically $V_1=V_2=0$. If, however, temperatures $T_1$ and $T_2$ take different values non-zero {\it thermoelectric} voltages
$V_{1,2}=V_{T1,2}$ are induced in our system. Introducing the renormalized subgap and
Sharvin resistances
\begin{gather}
\dfrac{1}{\tilde R_{1,2}}
=
\dfrac{1}{R_{1,2}}
-
\dfrac{2}{R^{nl}_+},
\quad
\dfrac{1}{\tilde R_{0}}
=
\dfrac{1}{R_{0}}
-
\dfrac{1}{R^{nl}_+},
\end{gather}
rewriting Eqs. \eqref{i1}, \eqref{i2} in the form
\begin{gather}
I_1 = \dfrac{V_{T1} - V_{T2}}{\tilde R_{0}}+\dfrac{V_{T1}}{\tilde R_1}
+ \tilde I^{nl},
\label{i11}
\\
I_2 = \dfrac{V_{T2} - V_{T1}}{\tilde R_{0}}+\dfrac{V_{T2}}{\tilde R_2} +
\tilde I^{nl}
\label{i22}
\end{gather}
and resolving these equations with respect to $V_{T1}$ and $V_{T2}$ together
with Eq. \eqref{discon}, we arrive at the result
\begin{equation}
V_{T1,2} =
\dfrac{
\tilde R_{2,1}(\tilde R_{1,2}+\tilde R_{0}/2)}{
\tilde R_1+ \tilde R_2+\tilde R_{0}}
\tilde I^{nl}(T_{1,2},V_{T1,2},\chi),
\label{vt12}
\end{equation}
which determines the magnitude of the thermoelectric voltages $V_{T1,2}$ induced by a nonzero temperature gradient $T_2-T_1$.

Eq. \eqref{vt12} -- together with the expression for $\tilde I^{nl}$ \eqref{tnl} -- constitutes the central result of this work.
It demonstrates that the thermoelectric effect in Andreev interferometers is in general not reduced by the small parameter $T/\varepsilon_F$ and
remains well in the measurable range. The key physical reason for this behavior is the presence of electron-hole
asymmetry generated for quasiparticles moving between two normal terminals and being scattered at two interfaces NS$_1$ and NS$_2$.
According to Eq. \eqref{as2} this asymmetry is generated for any value of the phase difference $\chi$ between superconductors S$_1$ and S$_2$
except for $\chi =0,\pi$ and for all values of the interface transmissions $\mathcal{D}_{1,2}$ except for $\mathcal{D}_{1,2}=0,1$.

These observations emphasize a crucial role played by coherent Andreev reflections at both NS interfaces.
Indeed, the effect trivially vanishes  $\tilde I^{nl}\equiv 0$ in the absence of Andreev reflection at any of the interfaces,
i.e. for $\mathcal{D}_{1(2)}=0$. Remarkably, electron-hole asymmetry is {\it not} generated also at full transmissions $\mathcal{R}_{1(2)}=0$
(cf. Eq. \eqref{as2}) and, hence, the current $\tilde I^{nl}$ \eqref{tnl} also vanishes provided complete Andreev reflection
($|\tilde \gamma_1^R|^2=1$) is realized at any of the NS interfaces. Moreover, bearing in mind that Andreev reflection does not violate
quasiparticle momentum conservation one can immediately conclude that, e.g., for  $\mathcal{R}_{1}=0$ electron scattering at the first
NS interface can only contribute to the BTK resistance of this interface but not to the non-local current $\tilde I^{nl}$.
This is because an incident electron and a reflected hole propagate along the
same (time-reversed) trajectories starting and ending
in one and the same normal terminal. This observation is specific for ballistic systems and has the same physical origin as,
e.g., the effect of vanishing crossed Andreev reflection contribution to the average non-local current in NSN structures with ballistic electrodes and
fully transparent NS interfaces \cite{KZ07}.

We would like to emphasize that the validity of our analysis is not restricted to the limit of small temperature
gradients and, hence, our results apply at any values $T_2-T_1$ provided both temperatures are kept well in the subgap range.
In the limit $T_2-T_1 \gg v_F/d$ the magnitudes of both thermal voltages $V_{T1,2}$ \eqref{vt12} and
the non-local current $\tilde I^{nl}$ \eqref{tnl} depend only on the lower of the two temperatures ($T_1$) and become practically
independent of the higher one ($T_2$). Then the maximum values of $\tilde I^{nl}$ which could possibly be reached at a given temperature
$T_1$ can roughly be estimated as
\begin{equation}
\tilde I^{nl}
\sim
\begin{cases}
e\mathcal{N}_{\mathrm{ch}}T_1 e^{-2\pi T_1d/v_F} , & T_1 \gg v_F/(2\pi d),
\\
 e\mathcal{N}_{\mathrm{ch}}v_F/d, &T_1 \ll v_F/(2\pi d),
\end{cases}
\label{estim}
\end{equation}
where $\mathcal{N}_{\mathrm{ch}} \sim p_F^2S$ is the number of conducting channels in a metallic wire with an effective cross section $S$. The parameter
$d$ here should be understood as an effective distance between the two NS interfaces. The estimate \eqref{estim} demonstrates that in the optimum case
the current $\tilde I^{nl}$ can be of the same order as the critical Josephson current of ballistic SNS junctions \cite{Kulik}
with similar parameters $\mathcal{N}_{\mathrm{ch}}$ and $v_F/d$. In the low 
temperature limit $T_{1,2} \ll v_F/(2\pi d)$ and small temperature difference 
$|T_1-T_2| \ll T_{1,2}$ Eqs. \eqref{tnl}, \eqref{W} yield $\tilde 
I^{nl} \propto (T_1+T_2)(T_1-T_2)$ in a qualitative agreement with the results \cite{Seviour00,KPV,VP,JW}.

Let us also note that the thermoelectric signal \eqref{tnl}, \eqref{vt12} 
derived within our model is described by an \textit{odd} periodic function of 
the phase difference $\chi$. This result goes in line with a general symmetry 
analysis \cite{VH2} both for diffusive and ballistic structures. At the same 
time, it is worth pointing out that odd as well as even $2\pi$-periodic phase 
dependencies of the thermopower were observed in experiments 
\cite{Venkat1,Venkat2,Petrashov03,Venkat3}. Possible explanations of such 
observations were proposed by a number of authors. E.g., Titov \cite{Titov} 
argued that an even in $\chi$ thermopower response can occur for certain 
geometries of Andreev interferometers as a result of the charge imbalance 
between the chemical potential of Cooper pairs in superconductor and the one for 
quasiparticles in the normal metal. Jacquod and Whitney \cite{JW} analyzed 
thermoelectric effects in Andreev interferometers within the scattering 
formalism and attributed an even in $\chi$ behavior of the observed thermopower 
to mesoscopic fluctuation effects.

\section{Josephson current and thermoflux}
\label{josephson}
In order to further investigate possible relation  between $\tilde I^{nl}$ and the Josephson current between the two S-terminals
let us evaluate the current $I_{12}$ flowing in the middle part of the normal wire, see Fig. \ref{nsnsn3-fig}. Under the condition
\eqref{discon} $I_{12}$ determines the total current between two superconductors S$_1$ and S$_2$. The whole calculation is carried out in the same manner as
that for the currents $I_1$ and $I_2$. Evaluating the contributions from all quasiparticle trajectories going through the central part
of the normal wire, we obtain
\begin{equation}
I_{12}=I_T+I_S^{\rm tot},
\end{equation}
where the term $I_T$ represents the thermoelectric current which has the form
\begin{equation}
I_T=\frac{V_{T1}}{R_{1}^+}
-\frac{V_{T2}}{R_{2}^-},\quad \frac{1}{R_{1,2}^{\pm}}=\frac{1}{R_{1,2}^*}+\dfrac{1}{R_{0}}\pm\dfrac{1}{R^{nl}_{-}},
\label{IT}
\end{equation}
$R_{1(2)}^*=R_{1(2)}|_{\mathcal{D}_{1(2)}=0}$, while the current $I_S^{\rm tot}$ defines the supercurrent between the two superconductors.
It can be split into two parts:
\begin{equation}
I_S^{\rm tot}=I_{S_1S_2}(\chi )+ I_S(\chi, T_{1,2},V_{T1,2}),
\end{equation}
where $I_{S_1S_2}(\chi )$ is the equilibrium Josephson current evaluated for closed quasiparticle trajectories confined between
the two S-terminals \cite{GZ02} (such trajectories, if exist, ``know nothing''
about the normal terminals $\mathrm{N_1}$ and $\mathrm{N_2}$ and
for this reason were not considered above) and
\begin{multline}
I_S(\chi, T_{1,2},V_{T1,2})=-e N_0
\int
\Biggr<16 \pi\bm{v}_{F}\Theta_{12}^{S_1S_2}(\bm{p}_F,\bm{r})
\\\times
 \dfrac{
\mathcal{R}_1\mathcal{R}_2\beta}{(\mathcal{R}_1 + \mathcal{R}_2)(1 + \mathcal{R}_1\mathcal{R}_2)}
\Biggl[
\frac{1+\mathcal{R}_2^2}{\mathcal{R}_2} T_1 W(\beta, t_1, \chi +v_1)
\\
+
\frac{1+\mathcal{R}_1^2}{\mathcal{R}_1} T_2 W(\beta, t_2, \chi -v_2)
\Biggr]
\Biggl>
d \bm{\Sigma}_1
\label{J}
\end{multline}
represents the contribution of open trajectories of the type $\Theta_{12}^{S_1S_2}$ connecting
the normal terminals $\mathrm{N_1}$ and $\mathrm{N_2}$. In Eq. \eqref{J} we
again defined $v_{1,2}=2eV_{T1,2} d/v_F$.

Comparing the above expressions for the thermoelectric current $I_T \propto \tilde I^{nl}$ and for the supercurrent $I_S^{\rm tot}$
we conclude that in general there exists no simple relation between these two currents. Only in the tunneling limit  $\mathcal{R}_{1,2} \to 1$
one can observe that Eqs. \eqref{tnl} and \eqref{J} are defined by almost the same integrals except for different signs
in front of the two $W$-functions in the square brackets. If, furthermore, we assume that the temperature difference exceeds the parameter
$v_F/d$, in the tunneling limit we obtain $\tilde I^{nl}\simeq I_S$ for $T_1-T_2 \gg v_F/d$ and $\tilde I^{nl}\simeq -I_S$ for $T_2-T_1 \gg v_F/d$.

Note that in the presence of a temperature gradient the expression \eqref{J} represents a {\it non-equilibrium} contribution to the 
Josephson current which explicitly depends on both $T_{1,2}$ and $V_{T1,2}$. In the absence of this gradient, i.e. at $T_1=T_2=T$ and $V_{T1,2}=0$
the term \eqref{J} reduces to its equilibrium form and we can define
\begin{equation}
I_J(\chi ,T)=-I_S(\chi, T_{1,2}=T,V_{T1,2}=0).
\label{IJ}
\end{equation}
Then in the tunneling limit $\mathcal{R}_{1,2} \to 1$ and provided both thermal voltages remain small $2eV_{T1,2} \ll v_F/d$ we may write
\begin{equation}
V_{T1,2}\propto \tilde I^{nl}\simeq \frac12[I_J(\chi ,T_1)-I_J(\chi ,T_2)].
\label{VJ}
\end{equation}
To a certain extent this relation resembles the result \cite{VH} derived in the diffusive limit.
Note, however, that within our model Eq. \eqref{VJ} is valid only under quite stringent conditions and, furthermore, the term $I_J(\chi ,T)$ \eqref{IJ} can be interpreted as a total equilibrium Josephson current only if we neglect the contribution $I_{S_1S_2}(\chi )$. Most importantly, it is clear from our analysis that the relation \eqref{VJ} by no means implies that large thermoelectric voltages $V_{T1,2}$ could possibly occur provided electron-hole symmetry in our system is maintained.

The effects discussed here can be conveniently measured, e.g., in a setup similar to that employed in recent experiments \cite{Petrashov16}. One can connect two superconductors in a way to form a superconducting loop.
Inserting an external flux $\Phi_x$ into this loop one would be able to control the phase difference $\chi =2\pi \Phi_x/\Phi_0$, where $\Phi_0$ is the superconducting flux quantum. By measuring the total flux inside
the loop $\Phi=\Phi_x+\mathcal{L}I_{12}$ with and without a temperature gradient one could easily determine
the value of the magnetic thermoflux
\begin{equation}
\Phi_T = \mathcal{L}[I_{12}(\chi, T_{1,2},V_{T1,2})-I_{12}(\chi, T)],
\end{equation}
where $\mathcal{L}$ is an effective inductance of a superconducting loop and $ I_{12}(\chi, T)=I_{12}(\chi, T_{1,2}=T,V_{T1,2}=0)$ is the equilibrium Josephson current at temperature $T$.

In conclusion, we demonstrated that large thermoelectric effect in multiterminal ballistic normal-superconducting hybrid structures is caused by electron-hole asymmetry generated for quasiparticles propagating between two normal terminals (kept at different temperatures) and suffering coherent Andreev reflection at two NS interfaces. At sufficiently high temperature gradients the thermoelectric voltages $V_{T1,2}$ depend only on the lowest of the two temperatures. The $2\pi$-periodic dependence of $V_{T1,2}$ on the superconducting phase difference $\chi$ is determined self-consistently and is strongly non-sinusoidal at low enough temperatures. Although the temperature dependence of $V_{T1,2}$ roughly resembles that of the equilibrium Josephson current in SNS junctions there exists no fundamental relation between these two quantities. Further information can be obtained by analyzing the behavior of the magnetic thermoflux induced in Andreev interferometers by applying
a thermal gradient.

This work was supported in part by RFBR grant No. 15-02-08273.

\appendix
\section{}
Let consider the integral
\begin{equation}
I=\int\limits_{-\infty}^{\infty}\dfrac{
\tanh\dfrac{\varepsilon+eV_2}{2T_2}
-
\tanh\dfrac{\varepsilon-eV_1}{2T_1}}{|a+be^{i\chi} e^{2i\varepsilon d/v_F}|^2}
d\varepsilon,
\end{equation}
where $a$ and $b$ are real positive numbers obeying the inequality $a>b$. This integral can be
conveniently evaluated by the method of complex contour integration. As a result we obtain
\begin{widetext}
\begin{multline}
I
=
\int\limits_{-\infty}^{\infty}
\Biggl[
-
\dfrac{b}{a^2-b^2}
\dfrac{e^{i\chi} e^{2i\varepsilon d/v_F}}{a+be^{i\chi} e^{2i\varepsilon d/v_F}}
-
\dfrac{b}{a^2-b^2}
\dfrac{e^{-i\chi} e^{-2i\varepsilon d/v_F}}{a+be^{-i\chi} e^{-2i\varepsilon d/v_F}}
+
\dfrac{1}{a^2-b^2}
\Biggr]
\left[
\tanh\dfrac{\varepsilon+eV_2}{2T_2}
-
\tanh\dfrac{\varepsilon-eV_1}{2T_1}
\right]
d\varepsilon
=\\=
\dfrac{2}{a^2-b^2}
\left(eV_1 + eV_2\right)
+
8\pi\dfrac{b}{a}
\dfrac{1}{a^2-b^2}
\Biggl[
T_2
W( b/a, t_2, \chi - v_2 )
-
T_1
W( b/a, t_1, \chi + v_1 )
\Biggr],
\end{multline}
where $t_{1,2} =2\pi T_{1,2} d/v_F$, $v_{1,2}=2eV_{1,2} d/v_F$ and the function $W(b/a, t, \chi)$ is defined in Eq. \eqref{W}.
\end{widetext}


\begin{thebibliography}{99}
\bibitem{NL} V.L. Ginzburg, Rev. Mod. Phys. \textbf{76}, 981 (2004).
\bibitem{Kalenkov12} M.S. Kalenkov, A.D. Zaikin, and L.S. Kuzmin, Phys. Rev.
Lett. \textbf{109}, 147004 (2012).
\bibitem{Machon} P. Machon, M. Eschrig, and W. Belzig, Phys. Rev. Lett. \textbf{110}, 047002 (2013).
\bibitem{Ozaeta}  A. Ozaeta, P. Virtanen, F.S. Bergeret, and T.T. Heikkil\"a,
Phys. Rev. Lett. \textbf{112}, 057001 (2014).
\bibitem{KZ14} M.S. Kalenkov and A.D. Zaikin,  Phys. Rev. B \textbf{90}, 134502 (2014).
\bibitem{KZ15} M.S. Kalenkov and A.D. Zaikin,  Phys. Rev. B \textbf{91}, 064504 (2015).
\bibitem{Beckmann} S. Kolenda, M.J. Wolf, and D. Beckmann, Phys. Rev. Lett. \textbf{116}, 097001 (2016).
\bibitem{Venkat1} J. Eom, C.-J. Chien, and V. Chandrasekhar, Phys. Rev. Lett. \textbf{81}, 437 (1998).
\bibitem{Venkat2} D.A. Dikin, S. Jung, and V. Chandrasekhar, Phys. Rev. B
\textbf{65}, 012511 (2001).
\bibitem{Petrashov03} A. Parsons, I.A. Sosnin, and V.T. Petrashov, Phys. Rev. B \textbf{67}, 140502(R) (2003).
\bibitem{Venkat3} P. Cadden-Zimansky, Z. Jiang, and V. Chandrasekhar, New J. Phys.
\textbf{9}, 116 (2007).
\bibitem{Seviour00} R. Seviour and A.F. Volkov, Phys. Rev. B \textbf{62}, R6116 (2000).
\bibitem{KPV} V.R. Kogan, V.V. Pavlovskii, and A.F. Volkov, EPL \textbf{59}, 875 (2002).
\bibitem{VH} P. Virtanen and T.T. Heikkil\"a, Phys. Rev. Lett. \textbf{92}, 177004 (2004); J. Low Temp. Phys. \textbf{136}, 401 (2004).
\bibitem{VP} A.F. Volkov and V.V. Pavlovskii, Phys. Rev. B \textbf{72}, 014529 (2005).
\bibitem{Eil} G. Eilenberger, Z. Phys. \textbf{214}, 195 (1968).
\bibitem{bel} W. Belzig, F. Wilhelm, C. Bruder, G. Sch\"on, and A.D. Zaikin,
Superlatt. Microstruct. \textbf{25}, 1251 (1999).
\bibitem{Schopohl95} N. Schopohl and K. Maki, Phys. Rev. B \textbf{52}, 490 (1995); N. Schopohl, cond-mat/9804064 (unpublished).
\bibitem{Eschrig00} M. Eschrig, Phys. Rev. B \textbf{61}, 9061 (2000).
\bibitem{Zaitsev84} A.V. Zaitsev, Sov. Phys. JETP \textbf{59}, 1015 (1984).
\bibitem{BTK} G.E. Blonder, M. Tinkham, and T.M. Klapwijk, Phys. Rev. B
\textbf{25}, 4515 (1982).
\bibitem{FN0} Quantum interference of time reversed trajectories describing 
multiple scattering of quasiparticles at boundaries of the normal wire may in 
general yield some modifications of the Andreev conductance (like, e.g., the 
so-called zero bias anomalies known to exist in the diffusive limit 
\cite{VZK,HN,Ben,Zai}). Such possible
    modifications, however, are not important for the effects discussed here.
\bibitem{VZK} A.F. Volkov, A.V. Zaitsev, and T.M. Klapwijk, Physica C {\bf 210},
21 (1993).
\bibitem{HN} F.W.J. Hekking and Yu.V. Nazarov, Phys. Rev. Lett. {\bf 71}, 1625
(1993); Phys. Rev. B {\bf 49}, 6847 (1994).
\bibitem{Ben} C.W.J. Beenakker, B. Rejaei, and J.A. Melsen, Phys. Rev. Lett.
{\bf 72}, 2470 (1994).
\bibitem{Zai} A.D. Zaikin, Physica B {\bf 203}, 255 (1994).
\bibitem{FN} Clearly, our results also include contributions from the tajectories $\Theta_{21}^{S_1}$, $\Theta_{21}^{S_2}$ and
$\Theta_{21}^{S_2S_1}$ which constitute time-reversed counterparts of respectively $\Theta_{12}^{S_1}$, $\Theta_{12}^{S_2}$ and $\Theta_{12}^{S_1S_2}$.
On the other hand, for simplicity here we ignore the contributions from trajectories of the type $\Theta_{12}^{S_2S_1}$ which can always
be made small by a proper choice of the system geometry.
\bibitem{KZ07} M.S. Kalenkov and A.D. Zaikin, Phys. Rev. B \textbf{76},
224506 (2007).
\bibitem{Kulik} I.O. Kulik, Sov. Phys. JETP \textbf{30}, 944 (1970); C.
Ishii, Progr. Theor. Phys. \textbf{44}, 1525 (1970).
\bibitem{JW} P. Jacquod, and R.S. Whitney, EPL \textbf{91}, 67009 (2010).
\bibitem{VH2} P. Virtanen and T.T. Heikkil\"a, Appl. Phys. A \textbf{89}, 625 (2007).
\bibitem{Titov} M. Titov, Phys. Rev. B \textbf{78}, 224521 (2008).
\bibitem{GZ02} A.V. Galaktionov and A.D. Zaikin, Phys. Rev. B \textbf{65},
184507 (2002).
\bibitem{Petrashov16} C.D. Shelly, E.A. Matrozova, and V.T. Petrashov, Science Advances \textbf{2}, 1501250 (2016).
\end{thebibliography}
\end{document}